\documentclass[aps,prb,twocolumn,showpacs]{revtex4}
\usepackage{graphicx}
\usepackage{dcolumn}
\usepackage{amsmath}
\usepackage{amssymb}

\begin{document}
\title{THE
ELECTRON-GAS PAIR DENSITIES AND THEIR NORMALIZATION SUM RULES IN TERMS OF
OVERHAUSER GEMINALS AND CORRESPONDING SCATTERING PHASE SHIFTS}
\author{Paul Ziesche}

\affiliation{Max-Planck-Institut f\"ur  Physik komplexer Systeme,
  N\"othnitzer Str. 38, D-01187 Dresden, Germany}
\date{\today}
\begin{abstract}
It is shown, how the normalization sum rules of the spin-parallel and 
spin-antiparallel pair densities of the homogeneous electron gas become sum 
rules for the scattering phase shifts of the Overhauser two-body wave functions
(geminals), with which 
the pair densities have been successfully parametrized recently. These new sum 
rules relate two-body quantities to a one-body quantity, namely the asymptotics
of the Overhauser geminals to the momentum distribution. \\
\end{abstract}


\newcommand{\gsim}{\mathrel{\raise.3ex\hbox{$>$\kern-.75em\lower1ex\hbox{$\=
sim$\
}}}}
\newcommand{\lsim}{\mathrel{\raise.3ex\hbox{$<$\kern-.75em\lower1ex\hbox{$\=
sim$\
}}}}

\maketitle


The (spin-unpolarized) homogeneous electron gas (HEG) is an important and 
widely used model for the phenomenon called electron correlation, cf. e.g. 
\cite{Mar}. Some of its details are hidden in the reduced densities. For the 
HEG ground state such characteristic quantum-kinematic quantities are (i) the 
momentum distribution 
$n(k)$ as the typical 1-body quantity (recently parametrized in terms of a 
special convex function \cite{Zie}) and (ii) the pair density (PD) $g(r)$ as 
the typical 2-body quantity (recently parametrized in Refs. \cite{Per}-
\cite{Gor2}). With 
spin-resolution one has to distinguish the PDs $g_{\uparrow \uparrow}(r)$ and
$g_{\uparrow \downarrow}(r)$ for electron pairs with an interelectron distance 
$r=|{\underline r}_1-{\underline r}_2|$ and with parallel and antiparallel 
spins, respectively. They describe the probabilities of finding in the 
neighborhood of an electron (the `blue' electron) another electron with the 
same respectively the opposite spin. So it holds 
$g_{\uparrow\uparrow}(r)\geq 0$ and $g_{\uparrow\downarrow}(r)\geq 0$. For 
large $r$ it is $g_{\uparrow\uparrow} (\infty)=1$ and 
$g_{\uparrow\downarrow} (\infty)=1$. For $r=0$ 
(`on-top') it is $g_{\uparrow\uparrow}(0)=0$ due to the Pauli repulsion and 
$g_{\uparrow\downarrow}(0)<1$ due to the Coulomb repulsion. These short-range 
deviations of 
$g_{\uparrow \uparrow}(r)$ and $g_{\uparrow \downarrow}(r)$ from their 
asymptotical values 1 are called Fermi hole and Coulomb hole, respectively. 
In between they show shell structures. They are normalized as (cf. e.g. 
\cite{ZieTheo}) 
\begin{equation}
\rho \int d^3r\, [1-g_{\uparrow\uparrow}(r)]=2, \quad 
\rho \int d^3r\, [1-g_{\uparrow\downarrow}(r)]=0
\label{1}
\end{equation}
with $\rho = $ electron density. The spin-summed PD $g(r)=
\frac{1}{2}[g_{\uparrow\uparrow}(r)+g_{\uparrow\downarrow}(r)]$ is therefore 
normalized as $\rho \int d^3r\, [1-g(r)]=1$, what is called perfect screening 
sum
rule or charge neutrality condition. Here and in the following wave lengths 
and momenta $k$ are measured in units of the Fermi wave length $k_{\rm F}= 
\frac{1}{\alpha r_s}, \; \alpha = (\frac{4}{9\pi})^{1/3}$ and lengths 
(like $r$ or $\rho ^{-1/3}$) in units of $k_{\rm F}^{-1}$. Therefore 
$\rho=\frac{1}{3\pi^2}$ and $\rho d^3r=\alpha^3d(r^3)$. The momentum 
distribution and the 
PDs depend parametrically on $r_s$, the radius of a sphere containing in the 
average one electron, measuring simultaneously the interaction strength with
$r_s=0$ corresponding to the ideal Fermi gas. 
In Refs. \cite{Per}-\cite{Gor2} the PDs as functions of the separation $r$ and
of the parameter $r_s$ have been fitted to the data of quantum Monte-Carlo 
calculations. With the help of (such available) PDs, particle-number 
fluctuations in fragments of the system can be discussed. \cite{Zie1} 
The partitioning into a generalized Hartree-Fock term and a connected remainder 
\begin{eqnarray}
g_{\uparrow\uparrow}(r) 
&=& g_{\uparrow\uparrow}^{\rm HF}(r)-h_{\uparrow\uparrow}(r), \quad 
g_{\uparrow\uparrow}^{\rm HF}(r) = 1-|f(r)|^2, 
\nonumber \\
g_{\uparrow\downarrow}(r) 
&=& g_{\uparrow\downarrow}^{\rm HF}(r)
-h_{\uparrow\downarrow}(r), \quad  
g_{\uparrow\downarrow}^{\rm HF}(r) = 1
\label{2}
\end{eqnarray}
defines the cumulant PDs $h_{\uparrow\uparrow}(r)$ and 
$h_{\uparrow\downarrow}(r)$. $f(r)=(2/N)\sum_{\underline k}n(k)
{\rm e}^{{\rm i}{\underline k}{\underline r}}$ is the dimensionless 1-body
reduced density matrix. $f(0)=1$ expresses the normalization of $n(k)$. The 
hole normalizations (1) imply
\begin{equation}
\rho \int d^3r\; h_{\uparrow\uparrow}(r)=2c, \quad 
\rho \int d^3r\; h_{\uparrow\downarrow}(r)=0,
\label{3}
\end{equation}
where $c=(2/N)\sum_{\underline k}n(k)[1-n(k)]$ is christened L\"owdin 
parameter \cite{Low} (measuring the correlation induced nonidempotency of 
$n(k)$, thus vanishing for $n(k)\to n^{(0)}=\Theta (1-k)$). It is easy to show
that 
\begin{equation}
g_{\uparrow\uparrow}^{\rm HF}(r)=2 g_-(r), \quad
g_{\uparrow\downarrow}^{\rm HF}(r)=g_+(r)+g_-(r),
\label{4}
\end{equation}
define a singlet term $g_+^{\rm HF}(r)$ and a triplet term $g_-^{\rm HF}(r)$ 
with
\begin{equation}
g_{\pm}^{\rm HF}(r)=\sum_L^{\pm}\nolimits <j_l^2(kr)>, \; L=(l,m_l),  
\label{5}
\end{equation}
where $\pm$ refer to even and odd $l$, respectively, and $\sum_L\cdots=\sum_l
(2l+1)\cdots$. The $k$-average is defined
as $<\cdots>= (2/N)\sum_{\underline k} \mu (k)\cdots$, 
$(2/N)\sum_{\underline k} \mu (k)=1$,  
$\mu(k)=(2/N)\sum_{\underline K} n(k_1)n(k_2)$, where 
$k_{1,2}=|\frac{1}{2}{\underline K}\pm {\underline k}|$. Note 
$(2/N)\sum_{\underline k}\cdots=\int_0^\infty d(k^3)\cdots $. Thus, 
$3 k^2 \mu (k)$ is the
probability of finding two electron momenta ${\underline k}_1$ and 
${\underline k}_2$ with the half difference $k$. Because of 
\begin{eqnarray}
\rho \int d^3r \left [g_{\pm}^{\rm HF}(r)-\frac{1}{2}\right ] &=& \pm (1-c), 
\nonumber \\
\rho \int d^3r \left [g_{\pm}(r)-\frac{1}{2}\right ] &=& \pm 1 , 
\label{6}
\end{eqnarray} 
the corresponding cumulant PDs $h_{\pm}(r)=g_{\pm}^{\rm HF}-g_{\pm}(r)$ are
normalized as
\begin{equation}
\rho \int d^3r\, h_\pm (r)=\mp c.
\label{7}
\end{equation}
$g_\pm(r)$ and $g_\pm^{\rm HF}(r)$ approach $1/2$ for $r\to \infty$, thus 
$h_\pm(\infty)=0$. Note $g(r)=[g_+(r)+3g_-(r)]/2$. \\

No correlation (corresponding to $r_s=0$) means $n(k)\to n^{(0)}(k)=
\Theta (1-k)$, thus $\mu(k)\to \mu^{(0)}(k)= 4\, (1-k)^2 (2+k)\Theta (1-k)$, 
cf. Ref. \cite{foo1}, and $h_\pm (r)=0$. In this ideal case the identities
\begin{eqnarray}
\sum_L^{\pm}\nolimits j_l^2(kr) &=& \frac{1}{2}
\left (1\pm\frac{\sin 2kr}{2kr}\right ), 
\nonumber \\
<\frac{\sin 2kr}{2kr}> &=& \left (\frac{3j_1(r)}{r}\right )^2,
\nonumber \\
\rho \int d^3r \, \left (\frac{3j_1(r)}{r}\right )^2 &=& 2
\label{8}
\end{eqnarray}
ensure the hole normalizations (1) to be obeyed. \\ 

Correlation (corresponding to $r_s\neq 0$) means $n(k)\neq \Theta (1-k)$ and
$h(r)\neq 0$. Recently the HEG-PDs for this case have been successfully 
parametrized in terms of Overhauser geminals (= 2-body wave functions) 
$R_l(r,k)$ \cite{Over}-\cite{foo}, namely as 
\begin{equation}
g_\pm (r)=\sum_L^\pm\nolimits <R_l^2(r,k)>.
\label{9}
\end{equation}
So the normalization (7) can be written as
\begin{equation}
\rho \int d^3r \sum_L^\pm\nolimits <[R_l^2(r,k)-j_l^2(kr)]>=\pm c 
\label{10}
\end{equation}
The question arises whether the lhs can be expressed in terms of quantities 
characterizing the differences between the Overhauser geminals $R_l(r,k)$ (for 
interacting electrons) and 
the Bessel functions $j_l(kr)$ (= Overhauser geminals for noninteracting 
electrons). Here it is shown that these characteristic quantities are the phase
shifts $\eta_l(k)$, which describe the large$-r$ asymptotics of the Overhauser 
geminals $R_l(r,k)$ according to
\begin{equation}
R_l(r,k)\rightarrow \frac{1}{kr}\sin\left (kr-l\frac{\pi}{2}+\eta_l(k)\right )
\quad {\rm for} \quad r\rightarrow \infty .
\label{11}
\end{equation}  
Namely, within the Overhauser approach a pair of electrons with momenta 
${\underline k}_1$ and
${\underline k}_2$ moving in the HEG is considered. Their center-of-mass
motion is decribed by ${\rm exp}({\rm i}{\underline K}{\underline R}
)$, where ${\underline K}={\underline k}_1+{\underline k}_2$ is 
the total momentum and ${\underline R}=({\underline r}_1+{\underline r}_2)/2$ is
the center-of mass coordinate. Their relative motion is described by 
$R_L({\underline r}, k)=R_l(r,k)Y_L({\underline e}_r)$ with the relative 
coordinate ${\underline r}={\underline r}_1-{\underline r}_2$ and the half
momenta difference $k=\frac{1}{2}|{\underline k}_1-{\underline k}_2|$. So the 
total 2-body wave function describing the pair is 
${\rm exp}({\rm i}{\underline K}{\underline R})
R_L({\underline r},k)$. The $R_l(r,k)$ are the solutions of the radial 
Schr\"odinger equation for the relative motion 
\begin{equation}
\left [-\frac{1}{r}\frac{\partial^2}{\partial r^2}r+\frac{l(l+1)}{r^2}+
v_{\rm eff}^{\pm}(r)-k^2\right ]
R_l(r,k)=0
\label{12}
\end{equation} 
with $v_{\rm eff}^\pm(r)=\frac{\alpha r_s}{r}+ \cdots $ as an effective 
interaction potential. The dots indicate the potential of the physically 
plausible screening cloud around each electron (resulting from exchange
and correlation), possibly different for "$+$" (even $l$) and "$-$" (odd $l$). 
Note that Eq. (\ref{12}) is the remainder of an effective 2-body Schr\"odinger 
equation and that $R_l(r,k)$ is (part of) a 2-body wave function, a so-called 
geminal.  Because the effective interaction potential $v_{\rm eff}^{\pm}(r)$ is
repulsive, the solutions of Eq. (\ref{12}) are
scattering states, characterized by the afore mentioned scattering phase shifts
$\eta_l(k)$. Therefore the integrals $\int d^3r \, j_l^2(kr)$ and 
$\int d^3r \, R_l^2(r,k)$ do not exist, but (the following proof is analog to 
that in Ref. \cite{Lan}) understanding the lhs's of Eq. (9) 
as $\int d^3r \cdots = \lim_{R \to \infty} \int_{r<R} d^3r \cdots$ and using
the identity
\begin{equation}
\frac{\partial}{\partial {\underline r}}
\frac{1}{2}\left ( 
\frac{\partial R_L^*}{\partial {\underline r}}
\frac{\partial R_L}{\partial (k^2)}
-R_L^*\frac{\partial^2 R_L}{\partial {\underline r} \partial (k^2)} +{\rm c.c.}
\right)
=R_L^*R_L
\label{13}
\end{equation}
(this follows from Eq. (\ref{12})) and the Gauss theorem
\begin{equation}
\int_{r<R} d^3r\, \frac{\partial}{\partial {\underline r}}
\left [ {\underline e}_r f(r)\right ]= 4\pi R^2 f(R),
\label{14}
\end{equation}
it turns out
\begin{eqnarray}
\int_{r<R}d^3r \, [R_l^2(r,k)-j_l^2(kr)]=
\label{15}
\end{eqnarray}
\begin{eqnarray}
\frac{4\pi R^2}{2k}\left (
\frac{\partial R_l}{\partial r}\frac{\partial R_l}{\partial k}
-R_l\frac{\partial ^2 R_l}{\partial r \partial k} 
-\frac{j_l}{\partial r}\frac{\partial j_l}{\partial k}
+j_l\frac{\partial^2 j_l}{\partial r \partial k}
\right )_{r=R}.
\nonumber
\end{eqnarray}
The rhs gives with Eq. (\ref{11}) the sum $A_l(k)+B_l(k)$, where
\begin{eqnarray}
A_l(k)=\frac{2\pi}{k^2}\eta'_l(k),
\label{16}
\end{eqnarray}
\begin{eqnarray}
B_l(k)=-\frac{\pi}{k^3}\left [ \sin \left ( 2kR-l\pi+2\eta_l(k) \right )-
\sin(2kR-l\pi) \right ].
\nonumber
\end{eqnarray}
These expressions have to be $k$-averaged (with $\mu(k)$) and $L$-summed. 
$\mu(k)$ has the properties $\mu(0)=8 (1-c)$, $\mu(\infty)=0$, $\mu'(k)<0$.
The discontinuity of $n(k)$ at $k=1$ (quasiparticle weight $z_{\rm F}$) makes 
$\mu''(k)$ also discontinous at $k=1$ and the correlation tail $n(k>1)\neq 0$ 
causes a corresponding correlation tail $\mu(k>1)\neq 0$. 
The phase shifts are gauged as $\eta_l(\infty)=0$. Because they originate
from a repulsive potential, it is $\eta_l(0)=0$. (For an attractive potential
the Levinson theorem would say $\eta_l(0)=n\pi$ with $n=$ number of bound
states in the $l-$channel, c.f. e.g. Ref. \cite{Gol} or \cite{Zie2},
p. 133.) In addition to this,
for repulsive potentials $v_{\rm eff}^\pm(r)$ with finite range it holds
$\eta_l(k)\sim k^{2l+1}$. The $k$-average $<\cdots>$ of $A_l(k)$ yields after
partial integration 
\begin{equation}
<A_l(k)>= 6\pi \int_0^\infty dk\, [-\mu'(k)] \eta_l(k)\, . 
\label{17}
\end{equation}
The other term gives $<B_l(k)>_1+<B_l(k)>_2$ with
\begin{eqnarray}
<B_l(k)>_1 &=& -\int_0^\infty dk\, b_{1}(k)\sin(2kR-l\pi), 
\nonumber \\
b_{1}(k) &=& \frac{3\pi}{k}\mu(k)[\cos 2\eta_l(k)-1], 
\nonumber \\
<B_l(k)>_2 &=& -\int_0^\infty dk\, b_{2}(k) \cos(2kR-l\pi),
\nonumber \\
b_{2}(k) &=& \frac{3\pi}{k}\mu(k)\sin 2\eta_l(k).
\label{18}
\end{eqnarray}
Successive partial integrations yield (semi-convergent) series starting with
\begin{eqnarray}
<B_l(k)>_1 = -\frac{1}{(2R)^3} [b_{1}''(1^-)-b_{1}''(1^+)]\cos (2R-l\pi)
\nonumber \\
+\frac{1}{(2R)^3} b_1''(0)\cos l\pi + \cdots ,
\nonumber 
\end{eqnarray}
\begin{eqnarray}
<B_l(k)>_2 = +\frac{1}{(2R)^2}b_2'(0)\cos l\pi +\cdots,
\label{19}
\end{eqnarray}
which vanish for $R\rightarrow \infty$. So the 2-body phase-shift sum rules
\begin{equation}
\frac{2}{\pi}\sum_L^\pm\nolimits \int_0^\infty dk\, [-\mu'(k)] \eta_l(k)=\pm c 
\label{20}
\end{equation}
follow. These new sum rules are conditions for the effective interaction 
potential 
$v_{\rm eff}^\pm(r)$, which generates the Overhauser geminals $R_l(r,k)$.  
Whether the phase shifts of the available Overhauser geminals 
\cite{Over}-\cite{Dav} really obey the sum rules (20) has to be checked. For 
$r_s=0$ (no correlation) the L\"owdin parameter $c$ and the Overhauser phase 
shifts $\eta_l(k)$ vanish. \\ 

The new sum rules (\ref{20}) are relations between the one-body momentum 
distribution $n(k)$ encoded in $\mu(k)$ and $c$ and the asymptotic behavior of 
the PD for large $r$ as described by the phase shifts $\eta_l(k)$ of the 
Overhauser two-body wave functions 
$R_l(r,k)$. This may be considered as complementary to the large-$k$ 
asymptotics of $n(k)$, which is determined by the on-top PD according to 
$n(k\to \infty)\sim g_{\uparrow\downarrow}(0)/k^8+O(1/k^{10})$. \cite{Kim,Yas}
A qualitative discussion of the mutual relation between $n(k)$ and $g(r)$ based
on the virial theorem is in Ref. \cite{Mac} \\     

Note, that the Overhauser geminals result from a 2-body Schr\"odinger 
equation with a local effective interaction potential $v^\pm_{\rm eff}(r)$, 
whereas the `true' geminals, namely
the natural geminals, which parametrize/diagonalize the 2-body reduced density
matrix, are presumably the solutions of a 2-body Schr\"odinger 
equation with a non-local effective interaction potential, similarly as the 
natural
orbitals, which diagonalize the 1-body reduced density matrix, follow from an 
1-body Schr\"odinger equation with a non-local effective 1-body potential. \\

Whereas Eq. (\ref{20}) is valid for a 2-body quantity of a uniform system 
(namely 
the HEG-PDs parametrized by scattering-state Overhauser geminals), the 
well-known Friedel sum rule for point defects in metals refer to a 1-body 
quantity of a non-uniform system (namely the charge distribution or screening 
cloud around the impurity). This cloud perfectly screens the impurity (in a 
spatially oscillatory manner - Friedel oscillations). The 
resulting impurity potential generates scattering states with phase shifts 
$\eta_l(k)$, such that the 1-body sum rule 
\begin{equation}
\frac{2}{\pi}\sum_L\nolimits \eta_l(k_{\rm F})=\Delta Z
\label{21}
\end{equation} 
holds. The lhs is the excess charge or the valence difference between the 
impurity and the host and the rhs is the total number of electronic states 
induced by the impurity. For the details of this subject cf. e.g. Refs. 
\cite{Zie2}, p. 167, 469 or \cite{Mer}, p. 44. For the generalisation of the 
Friedel sum rule to non-spherically symmetric scatterers cf. \cite{Joh}. For
the Sugiyama-Langreth neutrality sum rule of half-space jellium and its
generalisation cf. \cite{Lan} and \cite{Zie3}, respectively. \\

Note the essential difference between Eq. (\ref{21}), where an external charge 
causes a perturbation, and Eq. (\ref{20}), where the `blue' electron, which 
distorts 
its surrounding, is an internal charge belonging to the quantum many-body 
system with the consequences of (i) exchange (singlet/triplet or $\pm$ 
symmetry) and (ii) of the $k$ average $<\cdots>$. \\    

\section*{Acknowledgments}
The author thanks R. Asgari, P. Gori-Giorgi, K. Morawetz, and J. P.  Perdew 
for helpful discussions and acknowledges P. Fulde for supporting this work.


\begin{thebibliography}{99}
\frenchspacing


\bibitem{Mar}     P. Fulde, {\it Electron Correlation in Molecules and
                  Solids}, 3rd ed., Springer, Berlin, 1995;          
                  N. H. March, {\it Electron Correlation in Molecules and 
                  Condensed Phases}, Plenum, New York, 1996; N. H. March (ed.),
                  {\it Electron Correlation in the Solid State}, Imperial
                  College Press, London, 1999.

\bibitem{Zie}     P. Ziesche, phys. stat. sol. (b) {\bf 232}, 231 (2002);
                  P. Gori-Giorgi and P. Ziesche, Phys. Rev. B, Dec 15, 2002, in
                  press.

\bibitem{Per}   J. P. Perdew and Y. Wang, Phys. Rev. B {\bf 46}, 12947
                (1992); {\bf 56}, 7018 (1997).

\bibitem{Gor1}  P. Gori-Giorgi, F. Sacchetti, and G.B. Bachelet,
                Phys. Rev. B {\bf 61}, 7353 (2000), B {\bf 66}, 159901(E) 
                (2002).

\bibitem{Gor2}  P. Gori-Giorgi and J. P. Perdew, cond-mat/0206147 and Phys.
                Rev. B {\bf 66}, 165118 (2002).

\bibitem{ZieTheo} P. Ziesche, Solid State Commun. {\bf 82}, 597 (1992); 
                  P. Ziesche, J. Mol. Struc. (Theochem) {\bf 527}, 35 (2000), 
                  therein factors 1/2 are erroneously incorporated in the 
                  definition of $g_{\uparrow\uparrow}(r)$ 
                  and $g_{\uparrow\downarrow}(r)$, such that 
                  $g_{\uparrow\uparrow}(\infty)=g_{\uparrow\downarrow}(\infty)=
                  1/2$ instead of 1.

\bibitem{Zie1}  P. Ziesche, J. Tao, M. Seidl, and J. P. Perdew,
                Int. J. Quantum Chem. {\bf 77}, 819 (2000); P. Ziesche, in {\it
                Many-Electron Densities and Reduced Density Matrices}, 
                J. Cioslowski (ed.), Kluwer/Plenum, New York, 2000, p. 33. For 
                finite systems,
                particle number fluctuations in fragments have been discussed 
                by P. Fulde, cf. Ref. \cite{Mar}, p. 157; A. Savin, in {\it
                Review in Modern Quantum Chemistry}, K. D. Sen (ed.), World 
                Scientific, Singapore, 2002. 
                
\bibitem{Low}   P.-O. L\"owdin, Adv. Chem. Phys. {\bf 2}, 207 (1959).

\bibitem{Over}  A.~W.~Overhauser, Can. J. Phys. {\bf 73}, 683 (1995).

\bibitem{Gor3}  P. Gori-Giorgi and J. P. Perdew, Phys. Rev. B
                {\bf 64}, 155102 (2001).

\bibitem{Gor4}  P. Gori-Giorgi, in {\it Electron Correlations and Materials
                Properties} II, edited by A. Gonis, N. Kioussis, and 
                M. Ciftan, Kluwer/Plenum, New York, in press.

\bibitem{Dav}   B. Davoudi, M. Polini, R. Asgari, and M.P. Tosi, 
                cond-mat/0201423, and Phys. Rev. B {\bf 66}, 075110 (2002). In
                a subsequent
		paper (cond-mat/0206456) the same authors go beyond the 
                Hartree-like theory and show how the inclusion of exchange 
                and correlation through a Kohn-Sham equation let
                emerge liquid-like structures with increasing coupling
                strength $r_s$ through the formation of a first-neighbor shell
                and further oscillations in the PD $g(r)$, which for 
                larger $r_s$ becomes a precursor of the Wigner crystallisation. 

\bibitem{foo}   The success of the Overhauser approach seems to confirm the 
                possibility of a pair-density functional theory, the idea of 
                which is presented in Ref. \cite{Zie4}. Whereas therein the 
                geminal occupancy was erroneously assumed to be 1 and 0 
                according to occupied and 
                unoccupied geminals respectively (analog to the aufbau 
                principle of the Hartree-Fock and the Kohn-Sham schemes), in
                the Overhauser approach a geminal occupancy $\mu(k)$ is used 
                being quite different from a step function. For geminals there
                is no aufbau principle. 

\bibitem{Zie4}  P. Ziesche, Phys. Lett. A {\bf 195}, 213 (1994);
                Int. J. Quantum Chem. {\bf 60}, 1361 (1996);
                in {\it Electron Correlations and Materials Properties}, edited
                by A. Gonis, N. Kioussis, and M. Ciftan, Kluwer/Plenum, 
                New York, 1999, p. 361;
                M. Levy and P. Ziesche, J. Chem. Phys. {\bf 115}, 9110 (2001).

\bibitem{foo1}  The function $p(k)=3k^2\mu(k)$ appears also in the mentioned 
                fluctuation-correlation analysis of the HEG \cite{Zie1} and in
                Ref. \cite{Gill}.

\bibitem{Gill}  P. M. W. Gill, A. M. Lee, N. Nair, R. D. Adamson, J. Mol. 
                Struct. (Theochem) {\bf 506}, 303 (2002).

\bibitem{Gol}   M. L. Goldberger and K. W. Watson, Collision Theory, Wiley, New
                York, 1964 or 
                L. D. Landau and E. M. Lifshitz, {\it Quantum Mechanics}, 
                Pergamon, Oxford, 1977, \S 133. 

\bibitem{Zie2}  P. Ziesche and G. Lehmann (eds.),
                {\it Ergebnisse in der Elektronentheorie der Metalle},  
                Akademie/Springer, Berlin, 1983. 

\bibitem{Mer}   I. Mertig, E. Mrosan, and P. Ziesche, {\it Multiple Scattering 
                Theory of Point Defects in Metals: Electronic Properties }, 
                Teubner, Leipzig, 1987.

\bibitem{Joh}   W. John and P. Ziesche, phys. stat. sol. (b) {\bf 47}, 555 
                and K83 (1971); W. John, G. Lehmann, and P. Ziesche, phys. 
                stat. sol. (b) {\bf 53}, 287 (1972); P. Ziesche, J. Phys. 
                C{\bf 7}, 1085 (1974).

\bibitem{Lan}   D. C. Langreth, Phys. Rev. B{\bf 5}, 2842 (1972).

\bibitem{Zie3}  P. Ziesche and P. Gersdorf, phys. stat. sol. (b) {\bf 138}, 
                645 (1996). 

\bibitem{Kim}   J. C. Kimball, J. Phys. a {\bf 8}, 1513 (1975).

\bibitem{Yas}   H. Yasuhara and Y. Kawazoe, Physica A {\bf 85}, 416 (1976).

\bibitem{Mac}   W. Macke and P. Ziesche, Ann. Physik (Leipzig) {\bf 13}, 25 
                (1964).

\end{thebibliography}
\end{document}